\documentclass[letter]{jpsj2} 

\title{Thermal Symmetry Crossover and Universal Behaviors in Carbon
  Nanotube Dots}
 
\author{Haruka \textsc{Oguchi} and Nobuhiko \textsc{Taniguchi}}

\inst{Institute of Physics, University of Tsukuba, Tennodai
  Tsukuba 305-8571, Japan}

\date{\today}

\abst{
  Motivated by recent experiments on electronic transport through a
  carbon nanotube dot, we investigate the role of the intra- and
  inter-orbital Coulomb interactions on the temperature evolution of
  the conductance.  It is shown that small amount ($\lesssim 10\%$) of
  asymmetry between these Coulomb repulsions substantially deforms the
  conductance profile at finite temperature, particularly around
  half-filling. The nature of such thermal symmetry crossover is
  elucidated.
}

\kword{carbon nanotube, Coulomb interaction, quantum dot, Kondo
  effect, universality, Anderson impurity model}

\begin{document}
\maketitle%

In the last decade, electronic transport through semiconducting
nanostructures has been extensively studied for future applications of
quantum nano-devices.  As for a single quantum dot, we have now
understood fairly well that Coulomb interaction on the dot
fundamentally affects its transport.  It not only constrains electrons
to pass through a dot (the Coulomb blockade
phenomena~\cite{Kastner92} but also gives rise to the singular
conductance enhancement at low-temperatures in an odd electron number
on the dot (the Kondo effect~\cite{Goldhaber-Gordon98}).
A prominent feature of the latter phenomena is a universal temperature
dependence scaled by the Kondo temperature,
where low-temperature conductance quantitatively agrees with the
``universal curve'' predicted theoretically by the Anderson impurity
model~\cite{Goldhaber-Gordon98b}.

Recent experimental observations of the Kondo effect in carbon
nanotube (CNT) dots~\cite{Babic04,Jarillo-Herrero05,Makarovski07} have
rekindled interest in a role of strong interaction in two important
aspects: enlarged symmetry and apparently nonuniversal temperature
behavior.  While the standard Kondo effect occurs through a spin
one-half degenerate single level, the enlarged symmetry may augment
the role of the spin degree.  Carbon nanotube dots arguably have
almost doubly degenerate orbits in the topmost shell, producing a
variety of low-temperature
phenomena~\cite{Sasaki00,Oreg00,Nygard00,Izumida01,Borda03,Sasaki04,Galpin05,Galpin06}.
At a quarter filling $N_{d}\approx 1$, the entanglement between spin and
orbital degrees of freedom gives rise to the $SU(4)$ Kondo effect,
with the Kondo temperature one order of magnitude higher than the
standard case~\cite{Choi05}.
What is more intriguing is the temperature evolution around an even
valley $N_{d} \approx 2$. While the experiment by Makarovski
\textit{et al.}  has observed visible low-temperature ``Kondo-like
enhancement'' around $N_{d}=2$\cite{Makarovski07}, such enhancement
was absent in others~\cite{Babic04,Jarillo-Herrero05}.  Rather, a
characteristic `dip' is observed in the conductance profile around
half-filling ($V_{g}\sim 3 \textrm{V}$) (See Fig.~3(a) or
Supplementary Information Fig.~SI2 of Jarillo-Herrero \textit{et
  al.}~\cite{Jarillo-Herrero05} and compare Fig.~2~(a) of~Makarovski
\textit{et al.}~\cite{Makarovski07}).
A few possibilities of the nature at half-filling have been argued
actively so far, due to the singlet-triplet Kondo
effect~\cite{Jarillo-Herrero05}, the $SU(4)$ Kondo
effect~\cite{Makarovski07,Busser07,Anders08} with/without some more
complications.
Observed temperature dependence of conductance through CNT seems
nonuniversal rather than universal.

In this Letter, motivated by the mentioned experiments on
low-temperature transport through a CNT dot, we investigate the role of
the inter-orbital Coulomb repulsion and the thermal symmetry
crossover.  Our main concern is on the temperature dependence of the
linear conductance profile particularly around half-filling.
While the RG flow either at quarter-filling or at half-filling has been
shown to flow toward the $SU(4)$ symmetric strong coupling
point~\cite{Galpin06}, the absence of the exact $SU(4)$ symmetry may
affect substantially the behavior at finite temperature.
Indeed a small amount of interaction difference less than 10\% will
cause a large deformation to the conductance profile at finite
temperature (see Figs.~\ref{fig:small-dU} and~\ref{fig:large-dU}
later).  We point out that the thermal crossover by interaction
asymmetry provides a systematic understanding to observed thermal
evolutions~\cite{Babic04,Jarillo-Herrero05, Makarovski07}.


Relatively large energy scale of the Coulomb interaction allows us to
focus on the topmost electron shell. 
Regarding a CNT dot, a great success of Anders \textit{et al.}
explaining Makarovski \textit{et al.}  experiments makes us feel certain
that the $SU(4)$-symmetric model is a good starting point to model a
CNT dot system; the topmost shell of CNT doubly degenerates in orbits
$i=1,2$ and the view conforms to observations of $SU(4)$-Kondo effect
at quarter-filling~\cite{Choi05}.  The $SU(4)$-symmetric
model nevertheless misses something, failing to explain the
temperature evolution at half-filling of some
experiments~\cite{Babic04,Jarillo-Herrero05}.
Within the universality of the topmost shell, we regard the
interaction asymmetry among the orbits a clue. 

By this reasoning, we model the CNT dot by an orbitally degenerate
Anderson model with interaction asymmetry among orbits.
On the dot, an electron
interacts with an electron in the same orbit $i=1, 2$ by $U$ or in the
different orbit by $U'$. The total Hamiltonian is given by $ H = H_{D}
+ H_{L} + H_{T}$, where the dot $H_{D}$, the noninteracting leads
$H_{L}$, and the coupling between leads and the dot $H_{T}$ are
defined by
\begin{eqnarray}
  && H_{D} = \sum_{i\sigma} \left( \varepsilon_{d}\, \hat{n}_{i\sigma} + 
    U \hat{n}_{i\uparrow} \hat{n}_{i\downarrow} \right) + 
  U'\hat{n}_{1}\hat{n}_{2}, \\
  && H_{L} = \sum_{\alpha}\sum_{ki\sigma} \varepsilon_{k}\,
  c^{\dagger}_{\alpha ki\sigma}c_{\alpha ki\sigma},\\
  && H_{T} = \sum_{\alpha}\sum_{ki\sigma} \left(t_{k}
    c^{\dagger}_{\alpha k i \sigma}d_{i \sigma}+\text{h.c.} \right).
\end{eqnarray}
Here the number operator of the dot is defined by $\hat{n}_{i} =
\sum_{\sigma} \hat{n}_{i\sigma} = \sum_{\sigma} d^{\dagger}_{i\sigma}
d_{i\sigma}$ and the average electron number on the CNT dot $N_{d}
=\langle \sum_{i}\hat{n}_{i} \rangle$ is controlled from $0$ to $4$ by the gate
voltage $\varepsilon_{d}$.
In the calculation below, we assume constant density of states
$\rho_{\alpha}$ of the lead $\alpha$ and use
$\Gamma=\sum_{\alpha}\Gamma_{\alpha} =\sum_{\alpha} \pi \rho_\alpha
|t_k|^2$ as a coupling parameter between the leads and the dot.

In the case of $U'=U$, the total Hamiltonian $H$ retains the full
$SU(4)$ symmetry.  The states $(n_{1},n_{2})=(1,0)$ and $(0,1)$ are
degenerate at quarter-filling ($N_{d}=1$) and so are
$(n_{1},n_{2})=(2,0)$, $(1,1)$ and $(0,2)$ at half-filling
($N_{d}=2$).  When one breaks the $SU(4)$ symmetry by decreasing $U'$,
the degeneracy is broken at half-filling, but unbroken at
quarter-filling.  This simple argument indicates that the $SU(4)$
symmetry at half-filling is more vulnerable than that at
quarter-filling.  We will show later that this is indeed the case.

At $T=0$, the system is fully renormalized into the strong coupling
fixed point; the linear conductance is expected to depend only on
$N_{d}$, \textit{i.e.}, $G(T=0) =4G_{0} \sin^2 \big(\tfrac{\pi}{4}N_d
\big)$ (with $G_{0}= e^{2}/h$) according to the Friedel sum rule.
Such behavior has been confirmed in experiments~\cite{Makarovski07}.
Since $N_{d}$ changes only slightly by decreasing the asymmetric
parameter $\eta =U'/U$, the zero-temperature conductance is universal
and the effect of asymmetry $\eta <1$ emerges only at finite
temperature.

Our analytical approach is based on an extension of the
Kotliar-Ruckenstein formulation of slave-boson mean field theory
(KR-SBMT)~\cite{Kotliar86}, where a bosonic field is attached to each
type of local excitations rather than decoupling the charge and spin
degrees of freedom.  The approach has several advantages that other
slave-boson cousins miss: it retains finite Coulomb repulsion effect
and reproduces Fermi liquid behavior at $T=0$ with satisfying the
Friedel sum rule.  The KR formulation of SBMT is believed to be a
powerful non-perturbative method; it gives reliable results not only
qualitatively but also quantitatively up to the Kondo temperature,
agreeing successfully with numerical renormalization group methods and
experiments~\cite{Dong01,Takahashi06}.

When KR-SBMT is extended to the dot with doubly degenerate orbits, 16
bose fields are needed associated to each state of the dot: $e$ for
the empty, $p_{i\sigma}$ for one electron with orbit $i$ and spin
$\sigma$, $x_{i}$ for two electrons on the same orbit $i$,
$y_{s\sigma}$ for two electrons at different orbits with total spin
$(s,\sigma)$, $h_{i\sigma}$ for three electrons with a hole on
$i\sigma$, and $b$ for fully occupied state~\cite{Dong02,footnote1}.
To eliminate unphysical states, the completeness condition and the
correspondence condition between boson and fermion number
$n_{i\sigma}=Q_{i\sigma}=p^{\dagger} _{i\sigma}p_{i\sigma}
+x^{\dagger} _i x_i+y^{\dagger} _{1\sigma}
y_{1\sigma}+\frac{1}{2}(y^{\dagger} _{00} y_{00}+y^{\dagger} _{10}
y_{10}) + h^{\dagger} _{i \bar{\sigma}} h_{i
  \bar{\sigma}}+\sum_{\sigma'} h^{\dagger} _{\bar{i} \sigma'}
h_{\bar{i} \sigma'}+b^{\dagger}b$ are imposed in terms of Lagrange
multipliers $\lambda_{i\sigma}$.  By applying the mean field
approximation, all the boson fields are replaced by the expectation
values.  On determining these auxiliary parameters self-consistently
at each temperature and each gate voltage, the system reduces to the
renormalized resonant level model with the effective dot level
$\tilde{\varepsilon}_d$ and the effective hopping $\tilde{t}_{k}
=z_{i\sigma} t_{k}$,
\begin{equation}
  H_{\text{eff}} = H_{L} + \sum_{i} \tilde{\varepsilon}_d\,
  \hat{n}_{i} + \sum_{\alpha k i \sigma} \left( \tilde{t}_{k}\,
    c^{\dagger}_{\alpha k i\sigma}d_{i \sigma}+\text{h.c.} \right).
\label{eq:Heff}
\end{equation}
Here $z_{i\sigma} = (1-Q_{i\sigma}) ^{-\frac{1}{2}} [ e^{\dagger}
p_{i\sigma} + p^{\dagger}_{i \bar{\sigma}}x_i+p^{\dagger} _{\bar{i}
  \sigma}y_{1\sigma} + p^{\dagger}_{\bar{i} \bar{\sigma}}
(y_{00}+y_{10})/{2} + x^{\dagger}_{\bar{i}} h_{i \bar{\sigma}} +
(y^{\dagger}_{00} + y^{\dagger}_{10}) h_{\bar{i} \bar{\sigma}}/2 +
y^{\dagger}_{1\bar{\sigma}} h_{\bar{i} \sigma} + h^{\dagger}_{i\sigma}
b ] Q_{i\sigma}^{-\frac{1}{2}}$.  The effective Hamiltonian
eq.~\eqref{eq:Heff} conforms to the Fermi liquid
description~\cite{footnote2}. 
The form of $H_{\text{eff}}$ enables us to find the linear/nonlinear
conductance by the Meir-Wingreen formula~\cite{Meir92}.  We will
present the result of the linear conductance for the symmetric
coupling between the leads and the dot $\Gamma_{L}=\Gamma_{R}$ below.

The effective Hamiltonian eq.~\eqref{eq:Heff} defines a natural energy
scale, the characteristic temperature
\begin{equation}
  T^{*} = \sqrt{\tilde{\varepsilon}_d^{2} + \tilde{\Gamma}^{2}}\,
  \Big|_{T=0}\ .
\label{eq:T-star}
\end{equation}
where $\tilde{\Gamma}=\sum_{\alpha}\pi \rho_{\alpha}
|\tilde{t}_{k}|^2$ is the effective coupling between the dot and the
leads.  Interaction effect is encoded in terms of
$\tilde{\varepsilon}_d$ and $\tilde{\Gamma}$.  Note that the scale
$T^{*}$ is defined in the entire range of the gate voltage and it
reduces to the usual Kondo temperature at the Kondo valley $N_{d}=1$.
We evaluate renormalized parameters self-consistently at each
$\varepsilon_{d}$ and at temperature $T$.

\begin{figure}
  \begin{center}
    \includegraphics[width=0.9\linewidth]{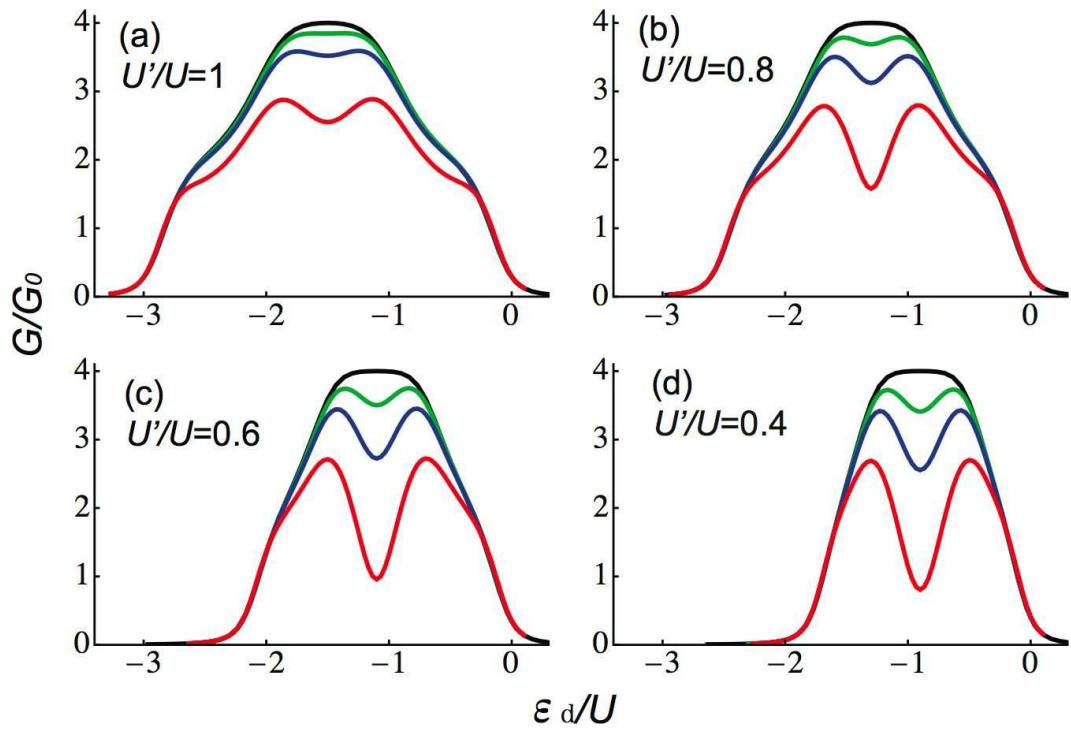}
  \end{center}
  \caption{(Color online) Temperature evolution of the conductance as
    a function of the gate voltage with varying asymmetry $U'/U=1.0$,
    $0.8$, $0.6$, and $0.4$.  Temperatures are $T/U=0$ (black),
    $1.67\times 10^{-3}$ (green), $3.33\times 10^{-3}$ (blue), and
    $8.33\times 10^{-3}$ (red).  $U$ is set to $U=30\Gamma$.}
  \label{fig:G-U-Uprime}
\end{figure}

Figure~\ref{fig:G-U-Uprime} demonstrates typical temperature
evolutions of the conductance profile as a function of the gate
voltage with increasing asymmetry (a)~$U'/U=1.0$, (b)~$0.8$,
(c)~$0.6$, and (d)~$0.4$, respectively.  Even with a small asymmetry
$U'/U=0.8$, a characteristic dip structure at half-filling develops
clearly at finite temperature, while only small deformation is seen
around quarter and three-quarter fillings.  Overall widths of
conductance peaks are determined by $U+2U'$.  Here the conductance profile
at $T=0$ can be considered universal, fully recovering
$SU(4)$ symmetry in the sense that it is determined by the Friedel sum
rule.  Interaction asymmetry manifests itself only at finite
temperature as a thermal crossover by modifying the conductance
profile substantially around half-filling.

\begin{figure}
  \begin{center}
    \includegraphics[width=0.9\linewidth]{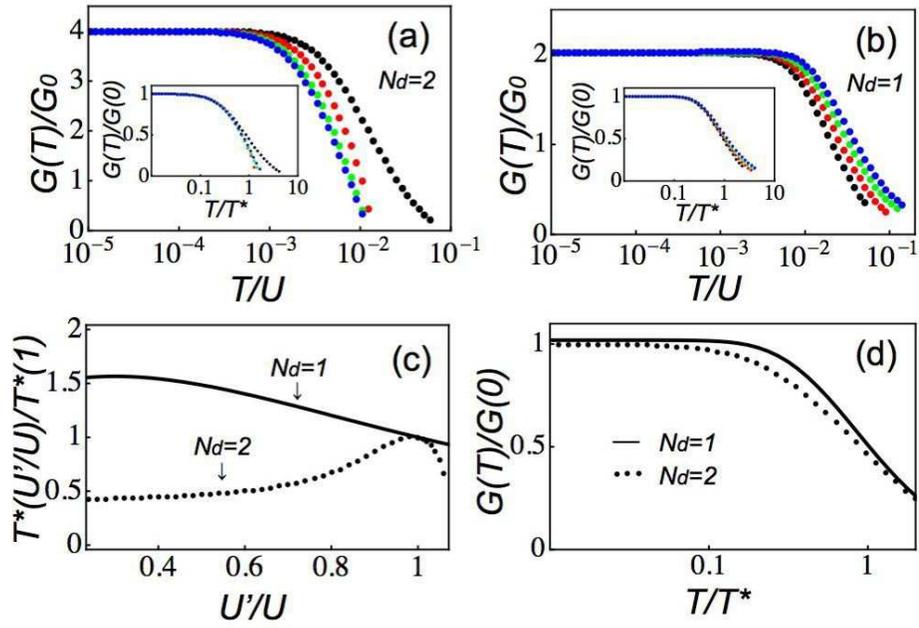}
  \end{center}
  \caption{(Color online) Temperature dependence of the conductance
    (a)~at $N_{d}=2$ and (b)~$N_{d}=1$ by varying $U'/U=1$ (black),
    $0.8$ (red), $0.6$ (green) and $0.4$ (blue).  Other parameters are
    the same with Fig.~\ref{fig:G-U-Uprime}. (c)~$T^{*}$ as a function
    of interaction asymmetry $U'/U$. (d)~Universal temperature dependence
   at $N_{d}=1$ (solid) and at $N_{d}=2$ (dotted) for $U'/U=1.0$.}
  \label{fig:compare-nd-one-two}
\end{figure}

\begin{figure}[t]
  \begin{center}
    \includegraphics[width=0.85\linewidth]{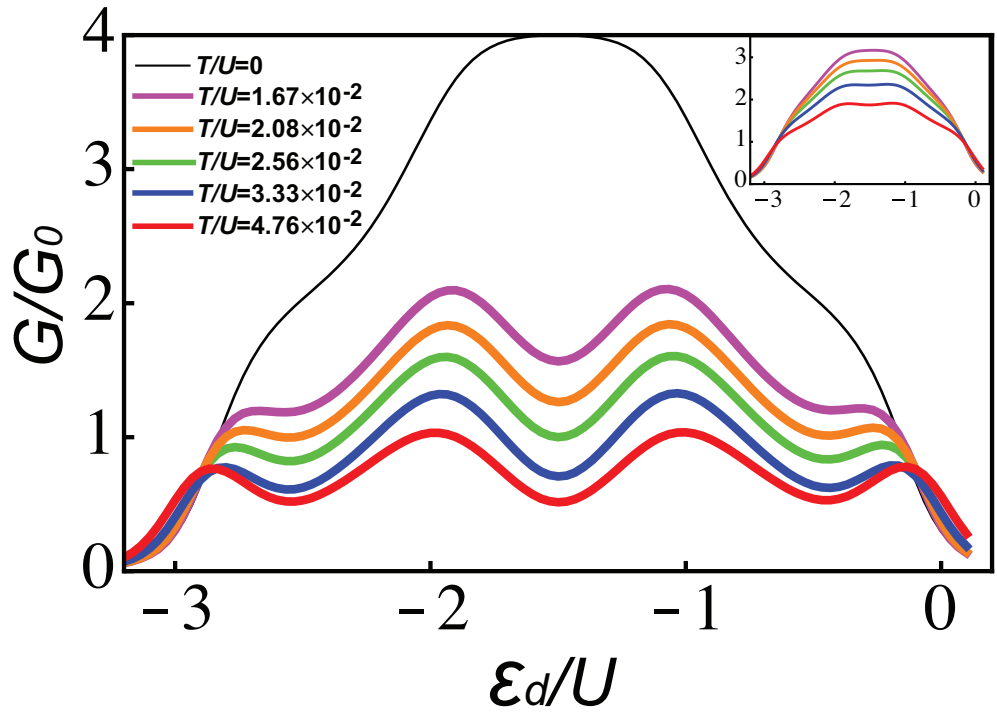}
  \end{center}
  \caption{(Color online) Schematic plotting for the comparison with CNT
    experiments~\cite{Makarovski07}: Conductance as a function of the
    gate voltage for $|U-U'| \ll T^{*}$, where $U=30\Gamma$ and
    $U'/U=0.997$.  Inset shows larger coupling parameter regime $U =
    15\Gamma$.}
  \label{fig:small-dU}
\end{figure}

\begin{figure}
  \begin{center}
    \includegraphics[width=0.85\linewidth]{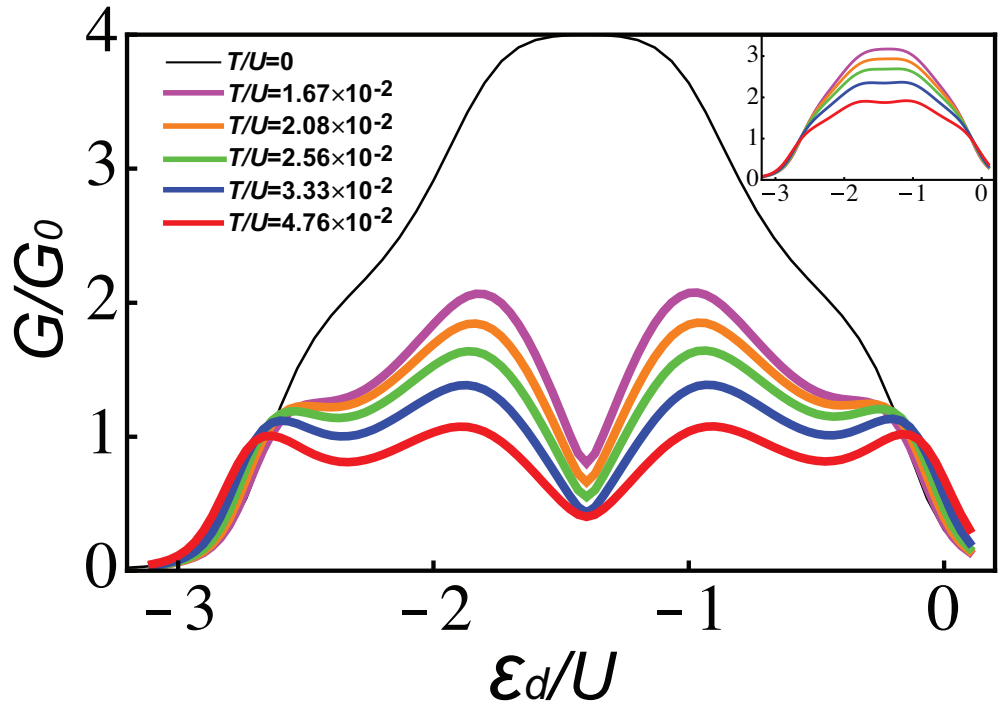}
  \end{center}
  \caption{(Color online) Schematic plotting for the comparison with CNT
    experiments~\cite{Jarillo-Herrero05}: Conductance as a function of
    the gate voltage for $|U-U'| \gg T^{*}$. Parameters are the same
    with Fig.~\ref{fig:small-dU} except for $U'/U=0.9$.}
  \label{fig:large-dU}
\end{figure}

To clarify the nature of the temperature dependence and understand
how the conductance is affected by the asymmetric
interaction particularly around $N_{d}\approx 2$, we make a direct
comparison between the behaviors at quarter-filling and half-filling.
Figure~\ref{fig:compare-nd-one-two} (a,b) shows the temperature
dependence of the conductance at (a) $N_{d}=2$ and (b) $N_{d} =1$ for
$U'/U=1$, $0.8$, $0.6$ and $0.4$.  By decreasing $U'$ away from
$U'=U$, it is clear that decreasing $U'$ reduces the characteristic
energy scale at $N_{d}=2$ but increases it at $N_{d}=1$.  The tendency
is elucidated by estimating the characteristic temperature $T^{*}$
defined in eq.~\eqref{eq:T-star}; while $T^{*}$ at $N_{d}=2$ is
already lower than that at $N_{d}=1$ at $U'=U$, the difference of
$T^{*}$ becomes more amplified by the presence of interaction
asymmetry (Fig.~\ref{fig:compare-nd-one-two}~(c)).  This behavior for
$N_{d}=2$ agrees well with other numerical
results~\cite{Galpin05,Busser07}.  
Clearly, the conductance enhancement around $N_{d}\approx 2$ is
destroyed more rapidly than at $N_{d}\approx 1$ at finite temperature,
which induces a substantially large deformation in the conductance
profile.
At the filling corresponding to Coulomb blockade
peaks ($N_{d}\approx 1.5, 2.5$), $T^{*}$ is found almost unchanged
by $U'/U$.

Another important observation is on the universal temperature
dependence.  Insets of Fig.~\ref{fig:compare-nd-one-two} (a,b) show
the data as a function of scaled temperature $T/T^{*}$.  As is seen,
the curves collapse well up to $T \lesssim T^{*}$ ($T^{*}$ is a upper
scale restricting the validity of the present analysis).  It implies
that the universal temperature dependence either at half-filling or at
quarter-filling reduces to the $SU(4)$ symmetric case of $U'=U$.
Additionally, Fig.~\ref{fig:compare-nd-one-two}~(d) shows that such
universal dependence differs slightly but significantly between
$N_{d}=1,2$~\cite{Anders08}.  It shows that the thermal symmetry
crossover observed at finite temperature is \emph{not} considered a
crossover between \emph{different universality classes} (between
$SU(4)$ and $SU(2)\times SU(2)$).  We can attribute the phenomena to
renormalizing the characteristic temperature $T^{*}$ by asymmetry
$U'/U$ at each gate voltage.

To stress the relevance of asymmetric interaction $U$ and $U'$ to
experimentally observed conductance
profiles~\cite{Babic04,Jarillo-Herrero05,Makarovski07}, we now present
schematic calculations mimicking experimental situation, changing
$U'/U$ slightly.  In the comparison, we have in mind a rough estimate
$U\approx 100\textrm{K}$, but we find the following characteristics
pretty generic.

Figure~\ref{fig:small-dU} demonstrates the conductance profile of
$U'/U=0.997$ with $U=30\Gamma$ ($U=15\Gamma$ for the inset).  With
this small amount of asymmetry, the profile reproduces all the
features of the $SU(4)$ symmetric model~\cite{Anders08}. A distinctive
feature of the conductance profile in large $U/\Gamma$ region is that
the Kondo enhancement by decreasing the temperature occurs either at
$N_{d}\approx 2$ or at $N_{d}\approx 1$ similarly.
Regarding a smaller value of $U/\Gamma$, four peaks merge to form one
big peak showing the Kondo enhancement (see the inset).  As was
claimed already~\cite{Anders08}, these temperature evolutions by 
the $SU(4)$ symmetric Anderson model agree very well with what is
observed either at $V_{g} \sim 3.9\textrm{V}$ or at $V_{g} \sim
5.3\textrm{V}$ in Makarovski \textit{et al.}~\cite{Makarovski07}.
When we resort to the (heuristic) prescription in choosing
$U/\Gamma$~\cite{footnote3}, our results agree very well to those by the
NRG results by Anders \textit{et al.}.  Detailed comparison will be
presented elsewhere.

Figure~\ref{fig:large-dU} illustrates the temperature evolution of the
conductance profile with all the same parameters with
Fig.~\ref{fig:small-dU} except for $U'/U=0.9$, whose value is deduced
from the Coulomb blockade peak spacings at
$T=8\textrm{K}$~\cite{Jarillo-Herrero05}.  The conductance profiles
regarding a smaller $U/\Gamma$ (the inset) are almost identical with
the symmetric case.  To our surprise, however, the conductance
profiles with four peaks ($U=30\Gamma$) is modified considerably by
this relatively small asymmetry (less than 10\%), having a
characteristic dip structure around half-filling.  It is noted that in
Fig.~3(a) or Fig.~SI2 of Jarillo-Herreo \textit{et
  al}.\cite{Jarillo-Herrero05}, the conductance at $V_{g} \approx
3\textrm{V}$ is smaller than that of $V_{g} \approx 2.8\textrm{V}$ or
$3.2\textrm{V}$ at each temperature.  We claim that
Fig.~\ref{fig:large-dU} captures well essential characteristics of
experiments~\cite{Babic04,Jarillo-Herrero05} with a reasonable
choice of parameters.

How much asymmetry brings the system away from the $SU(4)$ symmetric
behavior of Fig.~\ref{fig:small-dU} into that of
Fig.~\ref{fig:large-dU}?  The characteristic temperature $T^{*}$ at
half-filling (of the symmetric model) controls it; with $T^{*} = 1.4
\times 10^{-2}U$ in Figs.~\ref{fig:small-dU} and~\ref{fig:large-dU},
$|U-U'|\lesssim T^{*}$ is valid in the former but $|U-U'|\gg T^{*}$ in
the latter.  This difference affects the conductance profile at finite
temperature.

To understand experimental data fully, one more complication seems to
remain.  By decreasing the temperature below $8\textrm{K}$, one
conductance profile at half-filling ($V_g \approx 3\textrm{V}$) begins
to enhance, conforming to the Kondo-like effect, but it eventually
reduces below $2\textrm{K}$~\cite{Jarillo-Herrero05}.  
We regard such behavior beyond our scope of description with a
possible extra mechanism at an energy scale much smaller than the
Kondo temperature; some interesting possibilities such as a slight
difference of degenerate levels, exchange coupling and
orbital-dependent coupling have been argued, but it still remains to
be seen.


In conclusion, we have investigated the role of the inter- and
intra-orbital interactions in transport through a carbon nanotube dot.
By using the KR formulation of slave-boson mean field theory, we have
shown that a small amount of asymmetry between the intra-orbital and
inter-orbital interactions can give rise to a substantial effect on
the conductance profile at finite temperature by renormalizing
$T^{*}$.  It is suggested that interaction asymmetry at finite
temperature enables us to understand systematically the existing
experimental data~\cite{Babic04,Jarillo-Herrero05,Makarovski07}.  We
also anticipate to observe a similar crossover phenomena by applying
small amount of finite bias voltage because it should serve as an
energy cut-off similarly to finite temperature.

The authors appreciate W.~Izumida and H.~Tamura for helpful discussion.  The
work is partially supported by Grant-in-Aid for Scientific Research
(Grant No.~18500033) from the Ministry of Education, Culture, Sports,
Science and Technology of Japan.

\newpage

\newpage

\newpage

\newpage


\begin{thebibliography}{99}
\bibitem{footnote2} There exists an additional term of the
  (normal-ordered) renormalized interaction describing processes of
  two or more quasiparticle excitations relative to the interacting
  ground state.  Such processes, however, can be neglected here
  because our main interest is in the low-temperature transport.
\bibitem{footnote3} In the single-impurity Anderson model, it has been
  known~\cite{Takahashi06} that quantitative agreement can be attained
  between effective theories (the KR-SBMT or the functional RG) and the NRG
  method only if we choose the twice as large value of $U/\Gamma$ for the
  former. It is a heuristic prescription.
\end{thebibliography}

\end{document}